# Nature of the electromagnetic force between classical magnetic dipoles

Masud Mansuripur

College of Optical Sciences, The University of Arizona, Tucson

**Abstract**. The Lorentz force law of classical electrodynamics states that the force $F$ exerted by the magnetic induction $B$ on a particle of charge $q$ moving with velocity $V$ is given by $F = qV \times B$. Since this force is orthogonal to the direction of motion, the magnetic field is said to be incapable of performing mechanical work. Yet there is no denying that a permanent magnet can readily perform mechanical work by pushing/pulling on another permanent magnet — or by attracting pieces of magnetizable material such as scrap iron or iron filings. We explain this apparent contradiction by examining the magnetic Lorentz force acting on an Amperian current loop, which is the model for a magnetic dipole. We then extend the discussion by analyzing the Einstein-Laub model of magnetic dipoles in the presence of external magnetic fields.

**1. Introduction**. The force exerted by one magnetic dipole on another is capable of performing mechanical work, which appears to be in violation of the Lorentz force law of classical electrodynamics. The discrepancy is resolved if one examines the model of a magnetic dipole as an Amperian current loop, which allows for the balancing of the translational kinetic energy gained (or lost) by the moving current loop against its rotational kinetic energy. This problem will be analyzed in some detail in Sec.2, where we show that the change in the loop's rotational kinetic energy is also related to its changing magnetic dipole moment in accordance with Larmor's diamagnetic susceptibility of the loop.

We proceed to address, in Sec.3, the question of what would happen if a magnetic dipole failed to behave as an Amperian current loop. This brings up an alternative model of magnetism and magnetic dipoles, which will be treated in the context of the Einstein-Laub formulation of the classical theory. In Sec.4 we examine the electromagnetic (EM) energy of an immobile magnetic dipole in the presence of an externally-applied static magnetic field. Two approaches to EM energy, namely, the Lorentz approach (which considers magnetic dipoles as Amperian current loops), and the Einstein-Laub approach (which treats such dipoles as pairs of north-south magnetic monopoles) will be considered. Finally, in Sec.5, we extend the discussion of Sec.4 to the case of moving magnetic dipoles within a static magnetic field. The predictions of the Lorentz and Einstein-Laub formulations turn out to be in complete agreement in all the cases examined in Secs.4 and 5. The paper closes with a few concluding remarks in Sec.6.

**2. Interaction between two magnetic dipoles**. Consider a magnetic point-dipole $m_0 \hat{z}$, sitting at the origin of a spherical coordinate system $(r, \theta, \varphi)$, as shown in Fig.1. The magnetic field of the dipole in the surrounding free space is[1,2]

$$H(r, \theta, \varphi) = \frac{m_0}{4\pi\mu_0 r^3} (2\cos\theta \, \hat{r} + \sin\theta \, \hat{\theta}). \tag{1}$$

The reason for the free-space permeability $\mu_0$ appearing in Eq.(1) is that we are using the notation in which the magnetic induction is defined as $B = \mu_0 H + M$. According to this definition, the magnetization $M$ has the units of the $B$-field (i.e., weber/m$^2$ or tesla in $SI$, the international system of units), and the magnetic dipole moment of a small current loop of area $A$ carrying the current $I$ is $m = \mu_0 IA$.

Let a rigid circular ring of radius $R$ and negligible thickness be centered on the $z$-axis at $z = z_0$. The ring, which is uniformly charged around its circumference, has total charge $q$, total mass $m$, and rotates at a constant angular velocity $\Omega$ around the $z$-axis. The electric current circulating around the ring in the azimuthal direction $\hat{\varphi}$ is thus given by $I = q\Omega/(2\pi)$, the mechanical angular momentum of the ring is $\boldsymbol{L} = (mR^2\Omega)\hat{\boldsymbol{z}}$, and the magnetic dipole moment of the spinning ring is $\boldsymbol{m}_1 = \frac{1}{2}(\mu_0 qR^2\Omega)\hat{\boldsymbol{z}} = \frac{1}{2}(\mu_0 q/m)\boldsymbol{L}$.

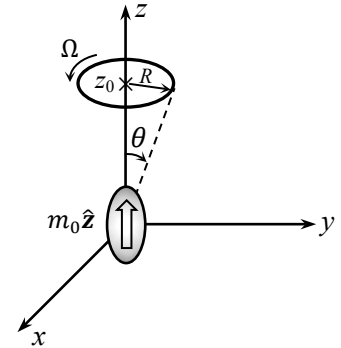

**Fig.1**. The magnetic dipole $m_0 \hat{z}$ sits at the origin of the coordinate system. A circular ring of radius $R$ and negligible thickness is placed parallel to the $xy$-plane at $z = z_0$. The ring, which is uniformly charged around its circumference, has a total charge $q$, total mass $m$, and rotates with a constant angular velocity $\Omega$ around the $z$-axis.

In the spherical coordinate system of Fig.1, the angle $\theta$ subtended by the ring at the origin is $\theta = \tan^{-1}(R/z_0)$. The Lorentz force-density (i.e., force per unit length of the ring) produced by the $B$-field of the dipole $m_0\hat{\boldsymbol{z}}$ acting on the spinning loop is readily found to be

$$\boldsymbol{f}_1 = I\hat{\boldsymbol{\varphi}} \times \frac{m_0(2\cos\theta\,\hat{\boldsymbol{r}} + \sin\theta\,\hat{\boldsymbol{\theta}})}{4\pi(R^2 + z_0^2)^{3/2}} = \frac{Im_0(2\cos\theta\,\hat{\boldsymbol{\theta}} - \sin\theta\,\hat{\boldsymbol{r}})}{4\pi(R^2 + z_0^2)^{3/2}}. \tag{2}$$

Upon integrating the above force-density around the perimeter of the ring, we find that only the $z$-component of the Lorentz force survives. The net force is thus given by

$$F_z = 2\pi R \frac{Im_0(-2\cos\theta\sin\theta - \sin\theta\cos\theta)}{4\pi(R^2 + z_0^2)^{3/2}} = -\frac{3Im_0 R\sin\theta\cos\theta}{2(R^2 + z_0^2)^{3/2}} = -\frac{3Im_0 R^2 z_0}{2(R^2 + z_0^2)^{5/2}}. \tag{3}$$

Incidentally, in the limit when $R \to 0$, the force $F_z\hat{\boldsymbol{z}}$ is equivalent to $(\boldsymbol{m}_1 \cdot \boldsymbol{\nabla})\boldsymbol{H}$, where $\boldsymbol{m}_1 = \mu_0\pi R^2 I\hat{\boldsymbol{z}}$ and $\boldsymbol{H} = m_0\hat{\boldsymbol{z}}/(2\pi\mu_0|z|^3)$ along the $z$-axis; see Eq.(1). The bottom line is that the dipole $m_0\hat{\boldsymbol{z}}$ exerts an attractive force $F_z\hat{\boldsymbol{z}}$ on the ring that is centered on the $z$-axis at $z = z_0$ while carrying the constant current $I$ along $\hat{\boldsymbol{\varphi}}$. When the radius $R$ of the ring is sufficiently small, one may suppose that the spinning ring is a magnetic point-dipole $m_1\hat{\boldsymbol{z}}$, in which case it is easy to see how the opposing poles of the two dipoles tend to attract each other. So long as the dipoles are pinned down at their respective locations, there will be no movement and, therefore, no mechanical work is performed by one dipole on the other. However, if the dipole at the origin remains fixed in place, but the dipole at $z = z_0$ is allowed to move along the $z$-axis, the force $F_z\hat{\boldsymbol{z}}$ of Eq.(3) will perform mechanical work, which appears to contradict the well-known assertion that magnetic fields acting on electric charges *cannot* do any work.

To resolve the contradiction, let us now assume that the ring spinning around the $z$-axis at $z = z_0$ is also travelling along $z$ at a constant velocity $v\hat{\boldsymbol{z}}$. The rotating charges are thus subjected to a new Lorentz force, whose linear density around the circumference of the ring is given by

$$\boldsymbol{f}_2 = \frac{qv\hat{\boldsymbol{z}}}{2\pi R} \times \frac{m_0(2\cos\theta\,\hat{\boldsymbol{r}} + \sin\theta\,\hat{\boldsymbol{\theta}})}{4\pi(R^2 + z_0^2)^{3/2}} = \frac{3qvm_0\sin\theta\cos\theta}{8\pi^2 R(R^2 + z_0^2)^{3/2}}\hat{\boldsymbol{\varphi}} = \frac{3qvm_0 z_0}{8\pi^2(R^2 + z_0^2)^{5/2}}\hat{\boldsymbol{\varphi}}. \tag{4}$$

The force-density $\boldsymbol{f}_2$ performs mechanical work $W$ on the rotating charges of the ring at a rate given by the dot-product of $\boldsymbol{f}_2$ and the azimuthal velocity $R\Omega\hat{\boldsymbol{\varphi}}$ of the charges. The rate at which work is done on the entire ring is subsequently obtained by integrating around the perimeter of the ring, that is,

$$\frac{dW}{dt} = \int_0^{2\pi}(\boldsymbol{f}_2 \cdot R\Omega\hat{\boldsymbol{\varphi}})R\,d\varphi = \frac{(3qvm_0 z_0 R\Omega)(2\pi R)}{8\pi^2(R^2 + z_0^2)^{5/2}} = \frac{3Im_0 R^2 z_0 v}{2(R^2 + z_0^2)^{5/2}}. \tag{5}$$

A comparison of Eq.(5) with Eq.(3) reveals that the energy delivered to the current loop in accordance with Eq.(5) is taken away from the translational kinetic energy of the traveling loop along the $z$-axis at the rate of $F_z v$. [Conversely, if $v$ happens to be negative, the loop gains translational kinetic energy as it moves down toward the dipole $m_0\hat{\boldsymbol{z}}$, while simultaneously losing rotational kinetic energy in accordance with Eq.(5).] We have thus confirmed that a magnetic field cannot perform any (net) work on moving charged particles — again for the simple reason that the Lorentz force of the $B$-field is always orthogonal to the direction of motion of each such charge.

The mechanical torque $(2\pi R^2 f_2)\hat{\boldsymbol{z}}$ acting on the ring is, of course, responsible for changing the ring's angular momentum $\boldsymbol{L}$ and, consequently, its dipole moment $\boldsymbol{m}_1 = \frac{1}{2}(\mu_0 q/m)\boldsymbol{L}$, at the following rate:

$$\frac{d\boldsymbol{m}_1}{dt} = \left(\frac{\mu_0 q}{2m}\right)\frac{d\boldsymbol{L}}{dt} = \left(\frac{\mu_0 q}{2m}\right)(2\pi R^2 f_2)\hat{\boldsymbol{z}} = \frac{3\mu_0 q^2 R^2 m_0 z_0 v}{8\pi m(R^2 + z_0^2)^{5/2}}\hat{\boldsymbol{z}} = \left(\frac{\mu_0 q^2 R^2}{4m}\right)\frac{3m_0 z_0 v}{2\pi(R^2 + z_0^2)^{5/2}}\hat{\boldsymbol{z}}. \tag{6}$$

The bracketed term on the right-hand-side of Eq.(6) is Larmor's diamagnetic susceptibility[3] associated with the magnetic dipole $m_1\hat{z}$. For sufficiently small $R$, the remaining term on the right-hand-side of Eq.(6), namely, $3m_0 v\hat{z}/(2\pi z_0^4)$, is the time-rate-of-change of the $B$-field as seen by the traveling point-dipole $m_1\hat{z}$ while passing through $z = z_0$ at the constant velocity $v\hat{z}$; see Eq.(1). It is seen that the dipole moment $m_1$ does in fact change in precisely the way it is expected to change (given its diamagnetic susceptibility) when passing through the nonuniform magnetic field produced by the stationary dipole $m_0\hat{z}$. The changing magnitude of the dipole moment $m_1\hat{z}$ guarantees that any gain (or loss) in its translational kinetic energy is balanced by the concomitant loss (or gain) in the rotational kinetic energy of the revolving charge(s) that constitute the dipole moment $m_1\hat{z}$.

**3. Beyond the Amperian current loop model.** The analysis of the preceding section applies to magnetic dipoles that originate in the orbital motion of electrons around atomic/molecular nuclei. It is not clear if spin-based magnetic dipoles would behave in similar fashion. In the Einstein-Laub formulation of EM force and torque,[4,5] magnetic dipoles are treated as primordial sources of the EM field (along with electric dipoles, in addition to free electric charges and currents). The Einstein-Laub force $F_{EL}$ and torque $T_{EL}$ exerted by the fields $E(r,t)$ and $H(r,t)$ on a magnetic dipole $m$ are[4-6]

$$F_{EL}(r,t) = (m \cdot \nabla)H - (\partial m/\partial t) \times \varepsilon_0 E, \qquad (7a)$$

$$T_{EL}(r,t) = r \times F_{EL} + m \times H. \qquad (7b)$$

Any changes that might occur in the dipole moment $m$ itself in the presence of external EM fields are unspecified. Moreover, with the Poynting vector defined as $S(r,t) = E(r,t) \times H(r,t)$ in the Einstein-Laub formulation,[5,6] the energy exchange rate between the EM field and a magnetic dipole $m$ is given by $H \cdot (\partial m/\partial t)$. A point-dipole such as $m_1\delta(x)\delta(y)\delta(z - vt)\hat{z}$, whose magnitude $m_1$ remains constant while traveling along the $z$-axis, will have the following energy-exchange-rate with the magnetic field $H(r)$ of Eq.(1):

$$\iiint_{-\infty}^{\infty} H \cdot (\partial m/\partial t)dxdydz = -\iiint_{-\infty}^{\infty} \frac{m_0 m_1 v}{2\pi\mu_0 z^3}\delta(x)\delta(y)\delta'(z - vt)dxdydz = -\frac{3m_0 m_1 v}{2\pi\mu_0 z_0^4}. \quad (8)$$

Equation (8) is in complete accord with Eq.(3), as it shows that translational kinetic energy is what the $H$-field delivers to (or takes away from) the moving dipole $m_1\hat{z}$. Unless the changes of $m_1$ with position along the $z$-axis are independently specified, the Einstein-Laub formalism cannot guarantee that the net work done by the $H$-field on the permanent dipole $m_1\hat{z}$ would be zero.

In contrast to the Einstein-Laub formalism, magnetic dipoles in the Lorentz formulation are treated as Amperian current loops,[1-3] with the current-density associated with an arbitrary spatio-temporal distribution of magnetization $M(r,t)$ being $J_{bound} = \mu_0^{-1}\nabla \times M$.[1,2,5] These current loops must then act as ordinary current loops in the presence of external EM fields. Given that, in their internal structure, magnetic dipoles (and also electric dipoles) do not abide by the rules of classical physics, it is by no means obvious if realistic (i.e., physical) dipoles are structurally capable of responding "properly" to classical EM fields. In other words, in the interaction between a magnetic field and magnetic dipoles, it is not a priori clear if the quantized spin and orbital angular momenta of realistic dipoles would allow the necessary adjustment of the magnetic dipole moment in order to ensure that no net mechanical work is performed by the applied magnetic field.

**4. Energy considerations for a stationary dipole.** With reference to Fig.2, let a small, uniformly-magnetized spherical particle of radius $R$ have a magnetization $M(r,t) = m(t)\delta(x)\delta(y)\delta(z)\hat{z}$, where the temporal variations of $m(t)$ are assumed to be fairly slow. The volume occupied by the spherical dipole is $V_0 = 4\pi R^3/3$. Here we are using the $\delta$-function notation to represent a small spherical volume centered at the origin of coordinates. In what follows, depending on the mathematical needs of the situation, we shall go back and forth between this $\delta$-function notation and the small-sphere representation of $M(r,t)$.

In the Lorentz formalism and in the absence of external fields, the time-rate-of-change of the stored EM energy inside the dipole, where $\boldsymbol{B}(t) = \frac{2}{3}\boldsymbol{M}(t)$, is given by

$$\frac{d\mathcal{E}_{\text{stored}}}{dt} = \frac{d}{dt}\left[\frac{1}{2}\mu_0^{-1}\int_{\text{sphere}}\boldsymbol{B}^2(t)d\boldsymbol{r}\right] = \frac{4}{9}\mu_0^{-1}V_0^{-1}m(t)m'(t). \quad \leftarrow \boxed{dr \text{ stands for volume element } dxdydz.} \quad (9)$$

The time-varying magnetization, being equivalent to a time-varying $B$-field inside the sphere, gives rise to an $E$-field circulating around the $z$-axis in accordance with Maxwell's equation $\boldsymbol{\nabla} \times \boldsymbol{E} = -\partial_t \boldsymbol{B}$. At the surface of the sphere, around a circle of radius $R\sin\theta$, we have $\oint \boldsymbol{E} \cdot d\boldsymbol{\ell} = -\frac{2}{3}M'(t)\pi R^2\sin^2\theta$. This $E$-field performs mechanical work on the magnet's bound electric current-density $\boldsymbol{J}_{\text{bound}}^{(e)} = \mu_0^{-1}\boldsymbol{\nabla} \times \boldsymbol{M}(t)$ and exchanges energy with the dipole at the rate of $\oint(\boldsymbol{E} \cdot \boldsymbol{J})d\boldsymbol{\ell}$. Given that the product of loop current and loop area integrated over $\theta$ equals $\mu_0^{-1}m(t)$, we see that the time-rate-of-change of the dipole's total internal energy is

$$\frac{d\mathcal{E}_{\text{total}}}{dt} = \frac{4}{9}\mu_0^{-1}V_0^{-1}m(t)m'(t) - \frac{2}{3}\mu_0^{-1}V_0^{-1}m(t)m'(t) = -\frac{2}{9}\mu_0^{-1}V_0^{-1}m(t)m'(t). \quad (10)$$

It is thus seen that, when $\boldsymbol{m}(t)$ is on the rise, electromagnetic (EM) energy leaves the interior of the dipole and enters the surrounding space in the form of EM fields. Conversely, when $\boldsymbol{m}(t)$ is in decline, the external field energy leaves the surrounding space and enters the dipole's spherical volume.

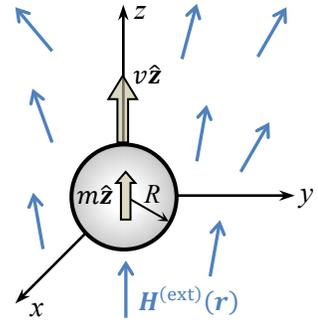

**Fig.2**. A uniformly magnetized sphere of radius $R$, volume $V_0$, and magnetic dipole moment $m(t)\hat{\boldsymbol{z}}$, moves along the $z$-axis at constant velocity $v\hat{\boldsymbol{z}}$. The internal $H$-field of the dipole is $-\frac{1}{3}\mu_0^{-1}V_0^{-1}m(t)\hat{\boldsymbol{z}}$, while its internal $B$-field is $\frac{2}{3}V_0^{-1}m(t)\hat{\boldsymbol{z}}$. The externally applied magnetic field $\boldsymbol{H}^{(\text{ext})}(\boldsymbol{r})$ does not vary with time.

Let us now switch to the Einstein-Laub formalism and confirm that the overall rate of flow of energy in or out of the sphere is the same in the two formulations. Here $\boldsymbol{H}(t) = -\frac{1}{3}\mu_0^{-1}\boldsymbol{M}(t)$ inside the sphere, and the time-rate-of-change of the energy stored within the dipole is given by

$$\frac{d\mathcal{E}_{\text{stored}}}{dt} = \frac{d}{dt}\left[\frac{1}{2}\mu_0\int_{\text{sphere}}\boldsymbol{H}^2(t)d\boldsymbol{r}\right] = \frac{1}{9}\mu_0^{-1}V_0^{-1}m(t)m'(t). \quad (11)$$

In this case, the $E$-field plays no role, but the exchange rate of energy-density between the field and the dipole is given by $\boldsymbol{H}(\boldsymbol{r}) \cdot \partial_t\boldsymbol{M}(\boldsymbol{r},t) = -\frac{1}{3}\mu_0^{-1}\boldsymbol{M}(\boldsymbol{r},t) \cdot \partial_t\boldsymbol{M}(\boldsymbol{r},t)$, which must then be integrated over the volume of the sphere. Thus, the time-rate-of-change of total (internal) energy is

$$\frac{d\mathcal{E}_{\text{total}}}{dt} = \frac{1}{9}\mu_0^{-1}V_0^{-1}m(t)m'(t) - \frac{1}{3}\mu_0^{-1}V_0^{-1}m(t)m'(t) = -\frac{2}{9}\mu_0^{-1}V_0^{-1}m(t)m'(t). \quad (12)$$

Note that Eqs.(10) and (12) predict identical rates for the overall flow of energy into or out of the dipole, despite the differing physical mechanisms behind the two formulations.

If the dipole happens to be immersed in an externally applied static magnetic field $\boldsymbol{H}^{(\text{ext})}(\boldsymbol{r})$, the results will be essentially the same. In this case, the Lorentz formulation predicts that

$$\frac{d\mathcal{E}_{\text{stored}}}{dt} = \frac{d}{dt}\left\{\frac{1}{2}\mu_0^{-1}\int_{\text{sphere}}\left[\mu_0\boldsymbol{H}^{(\text{ext})}(\boldsymbol{r}) + \frac{2}{3}\boldsymbol{M}(\boldsymbol{r},t)\right]^2d\boldsymbol{r}\right\}$$
$$= \frac{2}{3}m'(t)H_z^{(\text{ext})}(0) + \frac{4}{9}\mu_0^{-1}V_0^{-1}m(t)m'(t). \quad (13)$$

The external $H$-field will have no effect on the induced $E$-field and, therefore, the sole contribution of $\boldsymbol{H}^{(\text{ext})}$ to the overall time-rate-of-change of energy will be the addition of the first term on the right-hand side of Eq.(13), namely, $\frac{2}{3}m'(t)H_z^{(\text{ext})}(0)$, to $d\mathcal{E}_{\text{total}}/dt$ found in Eq.(10).

In contrast, in the Einstein-Laub formalism, when $\boldsymbol{H}^{(\text{ext})}(\boldsymbol{r})$ is present in the system, the time-rate-of-change of the energy stored inside the sphere will be

$$\frac{d\mathcal{E}_{\text{stored}}}{dt} = \frac{d}{dt}\left\{\tfrac{1}{2}\mu_0 \int_{\text{sphere}}\left[\boldsymbol{H}^{(\text{ext})}(\boldsymbol{r}) - \tfrac{1}{3}\mu_0^{-1}\boldsymbol{M}(\boldsymbol{r},t)\right]^2 d\boldsymbol{r}\right\}$$

$$= -\frac{1}{3}m'(t)H_z^{(\text{ext})}(0) + \frac{1}{9}\mu_0^{-1}V_0^{-1}m(t)m'(t). \tag{14}$$

This time, however, the rate of energy exchange between the $H$-field and the dipole becomes

$$\int_{\text{sphere}}\boldsymbol{H}(\boldsymbol{r},t)\cdot\partial_t\boldsymbol{M}(\boldsymbol{r},t)d\boldsymbol{r} = \int_{\text{sphere}}\left[\boldsymbol{H}^{(\text{ext})}(\boldsymbol{r}) - \tfrac{1}{3}\mu_0^{-1}\boldsymbol{M}(\boldsymbol{r},t)\right]\cdot\partial_t\boldsymbol{M}(\boldsymbol{r},t)d\boldsymbol{r}$$

$$= m'(t)H_z^{(\text{ext})}(0) - \tfrac{1}{3}\mu_0^{-1}V_0^{-1}m(t)m'(t). \tag{15}$$

Upon adding Eq.(14) to Eq.(15), we find

$$\frac{d\mathcal{E}_{\text{total}}}{dt} = \frac{2}{3}m'(t)H_z^{(\text{ext})}(0) - \frac{2}{9}\mu_0^{-1}V_0^{-1}m(t)m'(t). \tag{16}$$

It is seen, once again, that the sole contribution of the external field to $d\mathcal{E}_{\text{total}}/dt$ is the addition of the term $\tfrac{2}{3}m'(t)H_z^{(\text{ext})}(0)$ to the result obtained previously in Eq.(12). We conclude that the overall rate of EM energy flow into or out of a stationary dipole in a static magnetic field is independent of whether one works in the formalism of Lorentz or that of Einstein and Laub.

**5. Energy considerations in the case of a moving dipole**. A small spherical dipole of radius $R$ and volume $V_0$, immersed in the static external field $\boldsymbol{H}^{(\text{ext})}(\boldsymbol{r})$ and moving slowly with velocity $v\hat{\boldsymbol{z}}$ along the $z$-axis, can be represented by its magnetization $\boldsymbol{M}(\boldsymbol{r},t) = m(t)\delta(x)\delta(y)\delta(z - vt)\hat{\boldsymbol{z}}$; see Fig.2. In what follows, we will evaluate the integrals over a slightly expanded sphere of radius $R^+$ (denoted by sphere$^+$) in order to accommodate slight displacements of the dipole during an infinitesimal time interval $\Delta t$.

In the Lorentz formalism, the time-rate-of-change of the EM energy stored inside the (slightly enlarged) sphere is written

$$\frac{d\mathcal{E}_{\text{stored}}}{dt} = \frac{d}{dt}\left\{\tfrac{1}{2}\mu_0^{-1} \int_{\text{sphere}^+}\left[\mu_0\boldsymbol{H}^{(\text{ext})}(\boldsymbol{r}) + \tfrac{2}{3}\boldsymbol{M}(\boldsymbol{r},t)\right]^2 d\boldsymbol{r}\right\}$$

$$= \frac{2}{3}\int_{\text{sphere}^+}H_z^{(\text{ext})}(\boldsymbol{r})[m'(t)\delta(z - vt) - vm(t)\delta'(z - vt)]\delta(x)\delta(y)d\boldsymbol{r} + \frac{4}{9}\mu_0^{-1}V_0^{-1}m(t)m'(t)$$

$$= \frac{2}{3}m'(t)H_z^{(\text{ext})}(vt) + \frac{2}{3}vm(t)H_z'^{(\text{ext})}(vt) + \frac{4}{9}\mu_0^{-1}V_0^{-1}m(t)m'(t). \tag{17}$$

As for the contribution of the $E$-field induced by a time-varying $m(t)$, the exchanged energy will continue to have a time-rate given (to a good approximation) by $-\tfrac{2}{3}\mu_0^{-1}V_0^{-1}m(t)m'(t)$; see the paragraph immediately after Eq.(9). The time-rate-of-change of total energy within the dipole will thus be

$$\frac{d\mathcal{E}_{\text{total}}}{dt} = \frac{2}{3}m'(t)H_z^{(\text{ext})}(vt) + \frac{2}{3}vm(t)H_z'^{(\text{ext})}(vt) - \frac{2}{9}\mu_0^{-1}V_0^{-1}m(t)m'(t). \tag{18}$$

In contrast, in the Einstein-Laub formulation, we will have

$$\frac{d\mathcal{E}_{\text{stored}}}{dt} = \frac{d}{dt}\left\{\tfrac{1}{2}\mu_0 \int_{\text{sphere}^+}\left[\boldsymbol{H}^{(\text{ext})}(\boldsymbol{r}) - \tfrac{1}{3}\mu_0^{-1}\boldsymbol{M}(\boldsymbol{r},t)\right]^2 d\boldsymbol{r}\right\}$$

$$= -\frac{1}{3}\int_{\text{sphere}^+}H_z^{(\text{ext})}(\boldsymbol{r})[m'(t)\delta(z - vt) - vm(t)\delta'(z - vt)]\delta(x)\delta(y)d\boldsymbol{r} + \frac{1}{9}\mu_0^{-1}V_0^{-1}m(t)m'(t)$$

$$= -\frac{1}{3}m'(t)H_z^{(\text{ext})}(vt) - \frac{1}{3}vm(t)H_z'^{(\text{ext})}(vt) + \frac{1}{9}\mu_0^{-1}V_0^{-1}m(t)m'(t). \tag{19}$$

As for the exchange rate of EM energy between the field and the dipole, we write

$$\int_{\text{sphere}^+} \boldsymbol{H}(\boldsymbol{r},t) \cdot \partial_t \boldsymbol{M}(\boldsymbol{r},t) d\boldsymbol{r}$$

$$= \int_{\text{sphere}^+} [H_z^{(\text{ext})}(\boldsymbol{r}) - \tfrac{1}{3}\mu_0^{-1} M(\boldsymbol{r},t)][m'(t)\delta(z-vt) - vm(t)\delta'(z-vt)]\delta(x)\delta(y)d\boldsymbol{r}$$

$$= m'(t)H_z^{(\text{ext})}(vt) + vm(t)H_z'^{(\text{ext})}(vt) - \tfrac{1}{3}\mu_0^{-1}V_0^{-1}m(t)m'(t) - \tfrac{1}{3}\mu_0^{-1}vm(t)\partial_z M(\boldsymbol{r},t). \quad (20)$$

Thus the sum of Eqs.(19) and (20) agrees with the result obtained in the Lorentz formalism, Eq.(18), except for the last term on the right-hand side of Eq.(20), which appears to have no Lorentz counterpart. In fact, this last term is a mathematical artifact, as it represents the self-action of a fixed (i.e., time-invariant) dipole moving at constant velocity in free space, which must vanish. This last term also needs to be corrected, because the *H*-field of the dipole immediately outside its north and south poles needs to be included in the equation. The discontinuity of the spherical dipole's *H*-field at the sphere's surface requires that the self *H*-field at the surface be replaced by its average value across the discontinuity. The contributions of the last term in Eq.(20) to the integral, which are localized at the upper and lower hemispherical surfaces, thus cancel out.

As was the case in Sec.4 with regard to a stationary dipole, we conclude that the overall rate of EM energy flow into or out of a moving dipole in the presence of a static magnetic field does not depend on whether one works in the formalism of Lorentz or that of Einstein and Laub.

**6. Concluding remarks**. This paper started by asking the question: Why do magnets appear to perform mechanical work by pushing or pulling on other magnets (and also by attracting iron filings or pieces of scrap iron) when the Lorentz force law explicitly forbids such work? We argued that the answer depends on the model that one uses to represent magnetic dipoles. In the case of the Amperian current-loop model, the diamagnetic susceptibility of the dipole in the presence of an external magnetic field ensures that internal work (performed to speed up or slow down the current of the loop itself) cancels out the external work that is performed when the dipole is pushed or pulled by the external field. In the case of dipole as a north-south pair of adjacent magnetic monopoles (i.e., the Einstein-Laub model), the external work by the applied field is the same as that performed on an equivalent Amperian current loop. However, without additional information about the changes (if any) induced in the paired monopole type of dipole upon exposure to the external field, it is not possible to decide if the external work performed on the dipole is indeed cancelled by any internal work.

Both the Amperian current-loop model (in the context of what we called the Lorentz electrodynamics) and the model in which an adjacent pair of north-south monopoles comprise a magnetic dipole (i.e., in the Einstein-Laub formulation) can handle static as well as dynamic magnetic dipoles without violating Maxwell's equations or compromising the energy-momentum conservation laws; see Secs.4 and 5. The question remains as to which model can more accurately predict all the experimental observations.

**Acknowledgment**. The author is grateful to Ewan Wright and David Griffiths for commenting on an early version of the manuscript.